\documentclass{edm_article}

\usepackage{float}
\usepackage[moderate]{savetrees}
\usepackage{tabularx, colortbl}
\usepackage{listings}
\usepackage{booktabs}
\usepackage{rotating}
\usepackage{multirow}
\usepackage{adjustbox}
\usepackage{graphicx}
\usepackage[stable]{footmisc}
\usepackage{hyperref}
\usepackage{makecell}
\usepackage{balance}
\begin{document}
\definecolor{tablebar}{rgb}{0.6, 0.6, 0.9}
\setlength\arrayrulewidth{1pt}\arrayrulecolor{tablebar}


\title{Student Teamwork on Programming Projects\\ What can GitHub logs show us?
}

\numberofauthors{1}
\author{
Niki Gitinabard, Ruth Okoilu, Yiqao Xu, Sarah Heckman, Tiffany Barnes, \& Collin Lynch\\
       \affaddr{North Carolina State University}\\
       \email{\{ngitina, rookoilu, yxu35, sarah\_heckman, tmbarnes, cflynch\}@ncsu.edu}
}

 \maketitle
\begin{abstract}
Teamwork, often mediated by version control systems such as Git and Apache Subversion (SVN), is central to professional programming. As a consequence, many colleges are incorporating both collaboration and online development environments into their curricula even in introductory courses. In this research, we collected GitHub logs from two programming projects in two offerings of a CS2 Java programming course for computer science majors.  Students worked in pairs for both projects (one optional, the other mandatory) in each year. 
We used the students' GitHub history to classify the student teams into three groups, \textit{collaborative}, \textit{cooperative}, or \textit{solo-submit}, based on the division of labor.  We then calculated different metrics for students' teamwork including the total number and the average number of commits in different parts of the projects and used these metrics to predict the students' teamwork style. Our findings show that we can identify the students' teamwork style automatically from their submission logs. This work helps us to better understand novices' habits while using version control systems. These habits can identify the harmful working styles among them and might lead to the development of automatic scaffolds for teamwork and peer support in the future.
\end{abstract}



\section{Introduction}

Teamwork is an essential component of professional software development and CS educators incorporate it into their curricula to better prepare students for future careers \cite{feichtner84, reid05}. Working in teams provides students the opportunity to work on larger-scale projects than they otherwise would, and is more consistent with industry practice. Team projects also allow students to learn from their peers as described by \textit{Social Learning Theory} (SLT) \cite{bandura77}. \textit{Social Learning Theory} highlights four principal requirements for learning in social environments - \textbf{attention} or the opportunity to observe each other's work, \textbf{reproduction} or the chance to implement what they learned from observations, \textbf{retention} or being continuously engaged in the team, and \textbf{motivation} for learning \cite{bandura77}.

Prior research suggests that having all the team members engaged in the project is essential for success and for student learning. Seers et al. showed that the balance of contributions in a team is correlated with team performance and member satisfaction \cite{seers89}.
Chen et al. argued that uneven teamwork, where one member does the majority of the work, may limit the learning opportunities for their peers as well as themselves \cite{chen14}. Further, students that do not contribute may become less motivated to make genuine efforts since they can rely on their teammates to pick up the extra work \cite{van_der_Duim07}. 

Many researchers have measured and studied effective coding and study habits for individuals \cite{ahmadzadeh05, uchida02, spacco15, lin16,vihavainen13, carter15, blikstein11,watson14, hosseini14, chao16}. However, evaluating the quality of students' teamwork is more complicated. Hoegl et al. defined teamwork quality metrics as communication, coordination, balance of member contribution, mutual support, effort, and cohesion \cite{hoegl01}. Most of these metrics are not easy to quantify. Some have relied on surveys \cite{chen11assessing} or supervisor assessments \cite{nguyen16} for the evaluation of students' teamwork process, but little work has been done evaluating student teamwork quality and contribution in CS secondary education programming projects based on online activities \cite{parizi18}. As a result, in this work we aim to automatically identify the teams with weaker teamwork styles.

Teamwork in software development projects can be described in three ways. Coman et al. address two forms of teamwork: ``Collaborative'', where the teammates share the same goal toward solving an issue (in their case sharing a programming task), and ``Cooperative'', where they support each other while working on different goals \cite{coman14}. We observed that students took similar approaches in our group coding assignments. Some worked on similar parts of the project (e.g. both working on implementation or both testing, etc.) at the same time. Since each of these parts were focused on a specific goal (e.g. implementation adds program features and writing test cases improves code coverage and finds issues), working on the same parts means having a similar goal and these teams are similar to Collaborating teams. In such teams all the members have significant contributions to the same parts and they might even do pair-programming at times. Other groups of students divided the work by the project part (e.g. one works mostly on implementation and the other works on testing) while all contributing significantly. They might assist each other when necessary, but most of the work in each part was done by one member, focusing on the specific goal of each part. This behavior is similar to Cooperative work as mentioned by Coman et al.. The third form of teamwork is having a \textit{free-rider}, which as mentioned by van der Duim, is common in group projects \cite{van_der_Duim07}. We refer to this form of teamwork as ``Solo-submitting'' where one member did most of the work. The other members might have a few commits where they made a quick fix, but the majority of work was done by a single member. 

In this work, we analyzed data collected from two offerings of a CS2 course on Java programming for CS majors. 
In this course, students must complete three programming projects, one independent assignment, one where pairing is optional, and one where it is mandatory. The students are required to use GitHub as a version control system and for assignment submissions. Also, their GitHub repositories are connected to a Jenkins \cite{Jenkins} server, which provides them with responses from instructor-defined unit tests every time they submit their code.
We analyzed student commit behaviors to define metrics for evaluating their contributions to the team and to classify their work style. We tagged 400 commit messages as referring to different parts of the student projects (i.e. Implementation, Testing, Bugfix, Merge, Documentation, Style, and Other) that were graded in this course. We then used natural language processing to learn from that sample and tag the remaining commit messages. We finally used the commits in different categories to define several metrics for students' contributions to the team such as the \textit{Number of implementation commits} and the \textit{Percentage of testing}. In order to obtain a ground truth metric for the teamwork style we engaged two subject matter experts to classify 100 of the 238 team repositories in these classes into one of the three categories: ``Collaborative'', ``Cooperative'', and ``Solo-submit''. Then, we used the metrics to train and test prediction models on the teamwork styles of these projects. 

To be more specific, we test the following set of hypotheses:
\begin{itemize}
    \item H1. We can automatically classify commit messages into different parts of the project.
    \item H2. We can automatically classify student teams into Collaborative, Cooperative, or Solo-submit.
\end{itemize}

The findings of this study will help us to develop metrics to evaluate the effectiveness of student project teams and eventually provide students adaptive guidance or flag teams for instructor intervention. 

\section{Background}
Prior researchers have analyzed students' work on programming projects with the goal of identifying good habits that are common to higher performers and bad habits that are not \cite{ahmadzadeh05, uchida02, spacco15, lin16,vihavainen13, carter15, carter17, blikstein11,watson14, hosseini14, chao16}. Some researchers have also used visualization tools to analyze students' activity patterns and to present guidance to the students themselves (e.g. Retina \cite{murphy09, kim12}). One more recent approach to analyzing students' behavior is based upon studying logs from version control systems \cite{glassy06, reid05, murphy09, mierle05}. However, prior studies in this area have primarily focused on the students' individual work habits and not on the role that they play in a team. While other researchers have studied teamwork in CS courses (e.g \cite{naykki14, van06, barr05, oakley04, wen17, feichtner84, lee09}), these studies have generally relied upon student surveys and evaluations to bound their performance and only a few have considered their online behavior \cite{kim12, liu04, ganapathy11}. Thus, there is little prior work on detailed analyses of how individual student features affect team performance. 
While teamwork is the norm in industry, students may be unfamiliar with norms of collaborative work and many things can go wrong in team projects \cite{feichtner84}.  For example, some team members may decide to ``gang up'' and leave others out of the decisions or they might decide to be ``free-riders'' and do no work at all \cite{salomon89, van_der_Duim07}. A number of researchers have studied the impact of teamwork on student performance and ways to  enhance the experience of collaborative class work. Higher performing students often believe that they worked with greater initiative than their teammates, mostly alone, and they tend to give up on collaborative work \cite{lee17}. Additionally, there are users who prefer to work alone, mostly called ``lone wolves'' and their inclusion in teams often has a negative impact on the team's overall performance on the project \cite{barr05}. Instructors could use online contributions to easily identify some of these harmful patterns.

Another use for evaluating student contributions is to measure their teamwork quality. Most of the prior studies in classes evaluate the quality of the teamwork and their satisfaction with the teamwork experience based on the students' final peer evaluations \cite{van06,wen17,feichtner84,lee09,lee17}. While peer evaluation is a popular method among the instructors and is often used for grading group work \cite{oakley04, lee09}, it can be difficult to calculate student grades using their peers' estimations of their share of work \cite{feichtner84}. There are also other methods such as video-taping students while collaborating \cite{naykki14}. As suggested by Hoegl, the balance of students' contributions (i.e. having almost equal shares in the project) to the team is also an effective measure for team quality \cite{hoegl01}. Seers et al. also mentioned that the balance in the team members' contributions is related to team performance \cite{seers89}. However, measuring member contributions to software projects is not easy.

Other approaches have also been proposed for measuring team member contributions. One method relies on instructor qualitative evaluations \cite{main15, northrup06,imbrie05}. This opinion is often subjective and non-quantitative, but can provide good gestalt insights based on the students' online activities which makes the evaluation easier \cite{kim12, liu04, ganapathy11}. The same approach has also been used in software development projects in industry where the managers can view a summary of a team member's activities while evaluating their performance \cite{parizi18, nguyen16,lima15}. Kim et al. and Liu et al. suggested generating reports for the instructors based on version control system logs to track the students' activities and progress and intervene if needed \cite{kim12, liu04}. Such reports include information such as: who created the document, how many students edited the document, how many edits were made, how long the document was edited, how many words were included \cite{kim12}, total number of revisions, and the average number of work days \cite{liu04}. Studies have shown that these types of reports  can be used to track student team project progress and to intervene if necessary. 

While having the instructor or team manager's opinion is a reliable method to evaluate the contributions of different team members, Lima et al. noted that managers often find this evaluation time-consuming and that is has no specific criteria for good teamwork \cite{lima15}.  As a result, more recent studies have focused on automatically extracting the students' share of work from a version control system  \cite{ganapathy11}. For example, Ganapathy et al. evaluated group collaboration by the number of documents edited by several group members and found that better collaboration could predict a better outcome on the project \cite{ganapathy11}. El et al. similarly showed that the number of commits and the amount of lines of code added by a user are statistically significant characteristics for identifying contributions to the team.

\section{Dataset}
The dataset used in this study covers two consecutive fall semesters (2015 and 2016) of a CS2 Java programming course for majors. The course covers topics such as object-oriented design, testing, composition, inheritance, state machines, linear data structures, and recursion. Both course offerings were taught by the same instructors and were split into two on-campus sections. \hspace{1pt} All sections included two midterm exams (referred to as Test 1 and Test 2), a final exam, lab sessions, and three projects. The first project was completed individually while the students had the option to work in pairs for the second project, and were required to do so for the third. Students were allowed to request specific teammates or have them assigned by the instructor. When assigning students to teams, the instructor created balanced teams based upon similar prior performance on individual work (i.e. exam 1 and project 1). Both of the team projects included an individual component (a high-level system design and system test plan), and a team component (a system implementation). The system implementation part took about two weeks and the students were not permitted to work in a team if they failed to complete the individual task. Once students completed the individual parts and formed teams, an instructor-authored design was released and the students were required to implement it for the second stage of the project. 

\begin{table}
\centering
\caption{Statistics of Each Class}
\label{tab:stats}

\begin{tabular}{|l|c|c|} \hline
\textbf{Class} & \textbf{Java-2015} & \textbf{Java-2016}\\ \hline
On-campus Students & 181 & 206 \\
Teaching Assistants & 9 & 9 \\
On-campus Instructors & 2 & 1 \\
Average Grade & 79.7 & 79.9\\
Project 2 pairs & 36 & 44\\
Project 2 selected peers & 30 & 39\\
Project 2 assigned by instructor & 6 & 5\\
Project 3 pairs & 73 & 85\\
Project 3 selected peers & 39 & 56\\
Project 3 assigned by instructor & 34 & 29\\
Avg commits per repository & 109 & 66\\
Max commits per repository & 317 & 198\\
\hline
\end{tabular}
\end{table}

Students used the Eclipse IDE for the project implementation and were graded based on teaching staff and student-authored test cases, code coverage from student-authored test cases (EclEmma), coding style (SpotBugs, PMD, CheckStyle), and documentation (JavaDoc).
They used Moodle as a learning management system (LMS) to access materials and Piazza as a shared discussion forum. 
They also used the GitHub version control system to support teamwork and track coding progress. Whenever a student made changes to their project, a difference (diff) between the currently saved version and the edited version was created showing which files had been added or removed, and which lines of code had been added or removed. Students could store these diff changes by creating a ``commit'', which could serve as a checkpoint for progress. These commits were then uploaded, or ``pushed'', to GitHub, along with a commit message added by the students explaining the changes that had been made.

We focus our analysis on students' commit history as it reflects their coding behaviors. 

As shown in Table \ref{tab:stats}, the 2015 class had 182 students and 9 teaching assistants (TAs) while the 2016 class had 206 students and 10 TAs. Both these offerings included on-campus and distant education sections but we focused on the on-campus sections for consistency. In 2015, for the second part of projects, there were 39 pairs for Project 2 and 76 pairs for Project 3; the remaining students either failed to complete the design portion of the projects and worked alone or decided to work alone on project 2. In 2016, there were 46 pairs for Project 2, 88 pairs for Project 3, one group of three members for Project 2 and another group of three for Project 3, and the remaining students worked individually. Since the aim of our study is to understand the students' teamwork, we focused our analysis on Projects 2 and 3 and only on teams of 2 for consistency. 


%

\section{Methods}
\subsection{H1. We can automatically classify commit messages into different parts of the\\ project.}

Our dataset for 2015 contains a total of 4473 commits from Project 2 and 8224 from Project 3.  For 2016 we have a total of 7432 and 10430 commits for Projects 2 and 3 respectively. Since our focus is on the students' teamwork, we focused our analysis on commit messages of student pairs in Projects 2 and 3.

We first randomly selected and manually tagged 400 commit messages from our dataset classifying them into 7 different categories that described the commits. The tagging was done by a graduate student who had acted as a TA for this course several times before and was familiar with the structure of the projects. The categories were Implementation (I), Writing test cases (T), Bug fixing (B), Style fixing (S), Documentation (D), Merge (M), and Other (O). The distribution of these commit types among the 400 manually tagged commits and one example of each category are shown in Table \ref{commit_messages}. 

In 2016, students were taught about pair-programming and were specifically asked to mention it in their commit messages. We used keyword matching of words (e.g. ``pair'' as well as the whole word ``pp'') as some students abbreviated it to identify pair programming commits. We were able to find a total of 247 commits among all the student commits mentioning pair programming, 137 from Project 2 and 110 from Project 3, all in 2016 class. 

\begin{table}[]
    \centering
\begin{adjustbox}{width=0.48\textwidth}
    \begin{tabular}{|l|c|l|}
    \hline
 Commit Type & Percentage & Example \\
\hline
Implementation & 0.33 & Added Constructors for inner classes \\
Test Cases & 0.15 & More test cases \\
Bug Fixes & 0.29 & Fixed logout \\
Documentation & 0.03 & Added Javadoc to the class \\
Style & 0.04 & Fixing PMD errors \\
Merge & 0.03 & Merge branch 'master' of ... \\
Other & 0.11 & asdf \\
    \hline
    \end{tabular}
    \end{adjustbox}
    \caption{The distribution and an example of different commit types among manually tagged data}
    \label{commit_messages}
\end{table}

\begin{figure}[h]
    \centering
    \includegraphics[width=0.49\textwidth]{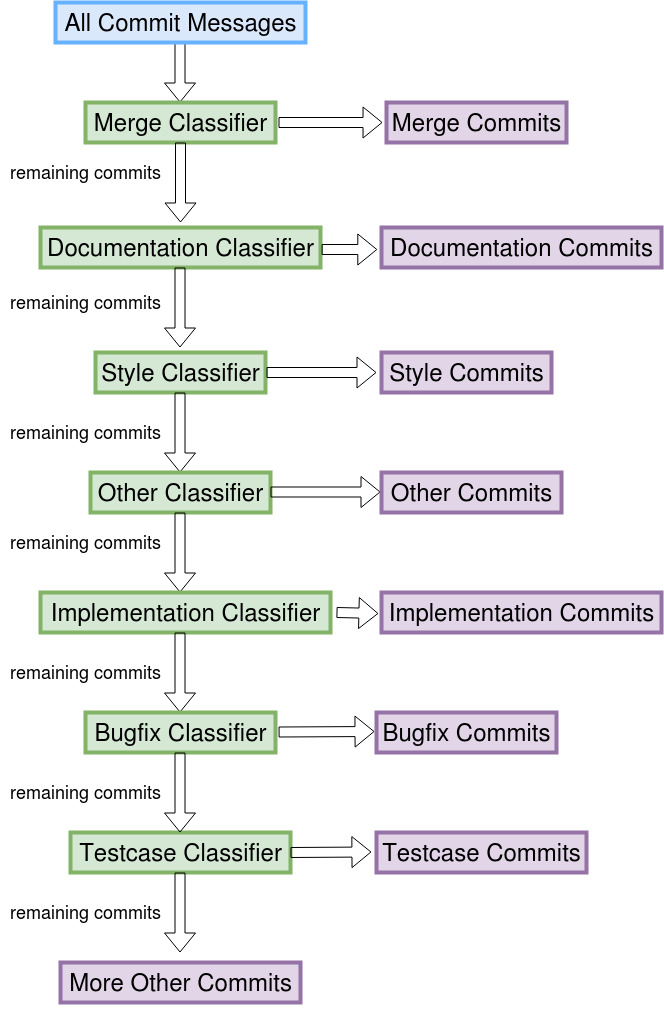}
    \label{commit_classification}
\end{figure}

For classifying the commit messages, we used a cascade model as shown in Figure \ref{commit_classification}. Some of these categories were easily identified by specific keywords. For example, merge commits are often auto-generated and always have the word ``merge'' in them, documentation commits often mention document or Javadoc keywords, style commits often mention the static analysis tools like PMD or CheckStyle, and commits that belong in none of our categories often do not have meaningful words and are easily detected using English corpus. We first removed English stop-words and lemmatized the text in the commit messages. We also added class-specific keywords to the acceptable English corpus, such as BBTP (black box test plan) or TS tests (teaching staff tests). To reduce the noise in our data and increase the accuracy of our models, we used static keyword matches to label \textit{merge}, \textit{documentation}, and \textit{style} and English corpus label \textit{other}. For the remaining tags (i.e. Implementation, Test cases, Bugfix), we used a Binary Logistic Regression classifier for each label, using TF-IDF vector of features \cite{tf-idf}, with a maximum of 45 features and an n-gram range of 1 to 4. Each binary classifier categorized a commit message as belonging to a category or not. Similar to the previous stage, we identified commits belonging in each category and removed the already-labeled commits from the dataset for the next prediction task. Any commits remaining unlabeled in the end were labeled as the Other category. After training and testing our classifiers on our tagged sample using 5-fold cross-validation, we used the trained models to predict the commit types for the remaining unlabeled commit messages.   

\subsection{H2. We can automatically classify student teams into Collaborative, Cooperative, or Solo-submit.}
We labeled the teamwork style of the students who both worked on similar parts of the project as \textit{collaborative} (e.g. both doing some implementation and some testing), while the teams where members both had significant contributions but mostly worked on separate parts of the project (e.g. one working on implementation and the other one on testing) were labeled \textit{cooperative}. There were also teams where one student did the majority of work and the other student either did no work, or made small amount of changes. We labeled those teams \textit{solo-submitting}. 

To identify the teamwork styles, we randomly selected 50 repositories from each offering of the course and manually tagged them as collaborative, cooperative, and solo-submit. The tagging was done by two subject matter experts (SME), experienced TAs who are familiar with the course material and grading criteria, one of whom has acted as a TA for this course multiple times. First, a sample of 20 repositories were tagged by both SMEs with a kappa agreement of 0.88 and then the remaining repositories were tagged separately. As mentioned before, the students were able to get feedback on their code by pushing it to GitHub and checking it with the teaching staff test cases. As a result, there are many cases where the students wanted to try different fixes and submitted many commits with small changes continuously until they could pass the tests. Thus, the SMEs were asked to focus on the amount of work done by each student in each category, rather than the number of commits. To make the tagging process more consistent, we added more specific definitions for the different teamwork styles. A team where both members contributed between 30\%-70\% to at least two common parts of the project were considered collaborative. The teams where one member did the majority of work in some parts and the other member worked mostly on other parts were considered cooperative. If one member did the majority of work in most parts and the other member did not work as much, the team was labeled as solo-submit.

Identifying teamwork style by manual tagging requires a great deal of expert time and it can be difficult to come to agreement among different experts. As a consequence, for this part we focused on extracting the students' teamwork automatically by using features from their GitHub submissions, as well as their prior individual performance (exam 1 and project 1) and the way they chose their team (i.e. self-selected vs. assigned by the instructor). In discussions, one of the instructors suggested that students with prior individual grades below 60 should be considered \textit{at-risk}.  Consequently, we added new binary features reflecting the team members' risk as well. Overall, we calculated the following features for each team member, sorting the team members such that the student with fewer total added lines of code in the project would be user 0 and the student with more would be user 1 in each team. 

Our final set of features included:
\begin{itemize}
    \item The \textbf{total number of commits} for each user in the whole project as well as the number of commits in each part (Implementation, Testing, Debugging, Documentation, Merge, Style, and Other). These features can show the students' contribution as the number of commits to the whole project and to the different parts.
    \item The \textbf{percentage of commits} for each user in the whole project and in different parts. This feature can distinguish between 2 commits in a team with a total number of 20 commits vs. in another team with a total of 100 commits. 
    \item The total number of \textbf{additions}, \textbf{deletions}, \hspace{1pt} \textbf{files changed}, and \textbf{amount of change} (i.e. additions + deletions) for each user in the whole project as well as each part. Additions and deletions in GitHub are measured by the lines of code each user changes in a specific commit.
    \item The \textbf{average} amount of \textbf{additions}, \textbf{deletions}, \textbf{files changed}, and \textbf{amount of change} per commit for each user in the whole project as well as each part.
    \item The \textbf{percentage} of each students' additions, deletions, files changes, and amount of change in the whole project as well as each part. Similar to the percentage of commits, this can normalize the amount of change for each team based on their total amount of activities. 
    \item The total and average \textbf{length of commit messages} for each user in the whole project and each part. This feature can distinguish between the members who write details about their changes and the ones who submit quick commits without much explanation.
    \item The total number of \textbf{pair programming commits} by each user as well as the total for the whole project. While using the total amount of pair programming is more intuitive, we believe that if all the pair programming is done on one person's computer, it might provide some information about the dynamics of the teamwork.
    \item Prior individual performance for each user (i.e. exam 1 and project 1 grades). Exam 1 and project 1 take place at a similar time and before project 2 and project 3.
    \item Risk label ($grade < 60$). We added each student's risk label for exam 1 and project 1 as separate features, as well as one overall risk label for the team which shows whether or not any member of the team could be considered at-risk based on exam 1 or project 1. 
    \item The team's selection method as a label ``selected'' which shows whether the students in this team requested working together or they were assigned by the instructor.
\end{itemize}

After defining and standardizing each feature, we ended up with 188 features. We used random forest feature selection as well as the recursive feature elimination (RFE) method with logistic regression to select the most important features for predicting students' teamwork style (i.e. collaborative, cooperative, or solo-submit). Random forests in Scikit-learn library return feature importance for all the features and we can select a desired number of top features for our model \cite{scikit-learn}. The RFE method in Scikit-learn library uses the coefficients of a linear model (in our case logistic regression) to estimate feature importance and prune the least important features until reaching the desired number of features \cite{scikit-learn}. We tried different numbers of features to find the features that resulted in better F1-scores. We then used cascade binary random forest and logistic regression classifiers using the selected features to predict each project's teamwork style. We chose these models because they are fast and they also provide us with information on what features they used and how those features contributed to the outcome, which can be useful when planning future interventions. Similar to commit classifications, these binary classifiers were trained based on belonging or not belonging to each category. We tested the accuracy of these models using 5-fold cross validation.


\section{Results and Discussion}
\subsection{H1. We can automatically classify commit messages into different parts of the\\ project.}

We first trained classifiers for the manually tagged commit messages in the categories of Style, Documentation, Merge, and Other using static matches. The F1-score, precision, and recall for these predictions are shown in Table \ref{SDMO_res}.
\begin{table}[]
    \centering
    \begin{tabular}{|l|c|c|c|c|}
    \hline
     & Merge & Style & Documentation & Other\\
     \hline
    F1 score & 0.98 & 0.99 & 0.99 & 0.95\\
    Precision & 1.00 & 1.00 & 1.00 & 1.00\\
    Recall & 0.96 & 0.98 & 0.99 & 0.90\\
    \hline
    \end{tabular}
    \caption{The performance of prediction models in finding Merge, Style, Documentation, and Other commits}
    \label{SDMO_res}
\end{table}

After removing these categories, we classified the remaining tagged commits into Implementation, Tests, and Bug fixes. In the end, any commits left in no category were categorized as Other. Since the list of Others commits changed after this stage, we calculated the accuracy of the models for this label twice, once based on the static analysis of the commit message as shown in Table \ref{SDMO_res}, and another time after assigning all the commits left uncategorized to this group as shown in Table \ref{ITBO_res}. The average F1-score, precision, and recall for the 5-fold cross validation for these predictive models are shown in Table \ref{ITBO_res}. As shown in this table, the precision of the Other category reduced as we added the remaining uncategorized samples to this group, which means some of these samples belonged to other categories but were not found by them, but the high recall score shows that all the commits in the Other category were identified successfully.  
\begin{table}[]
    \centering
    \begin{tabular}{|l|c|c|c|c|}
    \hline
 & Implementation & Bug Fix & Tests & Other\\
 \hline
F1 score & 0.84 & 0.92 & 0.92 & 0.78\\
Precision & 0.88 & 0.87 & 0.86 & 0.64\\
Recall & 0.82 & 0.97 & 0.98 & 1.00\\
    \hline
    \end{tabular}
    \caption{The performance of prediction models in finding Implementation, Bug Fixes, Tests, and Other commits}
    \label{ITBO_res}
\end{table}

These results support H1, showing that our prediction models are able to predict the categories of commit messages with an F1 score of 0.78 or higher. For most of the categories, the F1 score is higher than 0.9. After this step, we trained prediction models on all the tagged sample and used those models to predict the tags for the remaining untagged commit messages. The distribution of different commit messages in all the data is shown in Table \ref{labeled_commits}. These distributions show us that for all the projects and all the classes, a large portion of the students' commits belong to implementation and fixing bugs. Having very few style-based or documentation and tests commits shows that the students often fix style issues or add documentation and tests for the projects in fewer attempts. This is likely because they can check style errors and code coverage on their local platforms and submit once done, while adding features to their code and getting a functional version of the project that passes all the teaching staff test cases is often challenging and takes many attempts. Teaching staff tests were hidden from students and feedback was only available by committing code to GitHub that was then automatically executed on Jenkins. Students likely made frequent changes to address teaching staff test failures.

\begin{table}[]
    \centering
    \begin{adjustbox}{width=0.48\textwidth}
    \begin{tabular}{|l|cc|cc|cc|cc|}
    \hline
  & \multicolumn{4}{c|}{\textbf{2015}}& \multicolumn{4}{c|}{\textbf{2016}}\\
  \cline{2-9}
 & \multicolumn{2}{c|}{\textbf{Project 2}} & \multicolumn{2}{c|}{\textbf{Project 3}} & \multicolumn{2}{c|}{\textbf{Project 2}} & \multicolumn{2}{c|}{\textbf{Project 3}} \\
 & Count & Ratio & Count & Ratio & Count & Ratio & Count & Ratio \\
 \hline
 
Implementation & 848 & 0.25 & 2060 & 0.33 & 2666 & 0.44 & 4317 & 0.49 \\
Test Cases & 298 & 0.09 & 327 & 0.05 & 637 & 0.10 & 450 & 0.05 \\
Bug Fixes & 767 & 0.22 & 1433 & 0.23 & 1262 & 0.21 & 2076 & 0.24 \\
Documentation & 117 & 0.03 & 196 & 0.03 & 464 & 0.08 & 471 & 0.05 \\
Style & 141 & 0.04 & 268 & 0.04 & 457 & 0.08 & 624 & 0.07 \\
Merge & 367 & 0.11 & 520 & 0.08 & 173 & 0.03 & 533 & 0.06 \\
Other & 901 & 0.26 & 1399 & 0.23 & 433 & 0.07 & 340 & 0.04 \\
\hline
    \end{tabular}
    \end{adjustbox}
    \caption{The distribution of different commit types in each year and project}
    \label{labeled_commits}
\end{table}

\vspace{0.5cm}
\subsection{H2. We can automatically classify student teams into Collaborative, Cooperative, or Solo-submit.}


\begin{table}[]
    \centering
    \begin{adjustbox}{width=0.5\textwidth}
    \begin{tabular}{|l|cc|cc|cc|}
    \hline
        & \multicolumn{2}{c|}{\textbf{Total}} & \multicolumn{2}{c|}{\textbf{2015}} & \multicolumn{2}{c|}{\textbf{2016}}\\
 & Count & Ratio & Count & Ratio & Count  & Ratio \\
 \hline
\textbf{SME tagged} &  &  &  &  &  &  \\
Collaborative & 55 & 0.57 & 18 & 0.39 & 37 & 0.76 \\
Cooperative & 28 & 0.29 & 18 & 0.39 & 10 & 0.20 \\
Solo-submitting & 12 & 0.14 & 10 & 0.22 & 2 & 0.04 \\

        \hline
    \end{tabular}
    \end{adjustbox}
    \caption{Distribution of the different teamwork styles}
    \label{repo_distribution}
\end{table}

In our SME-tagged data, we identified a total of 14 solo working teams, 55 collaborating teams and 28 cooperating teams. We removed five teams from our analysis that had more than two members or only one member contributing either because they were teams of 3 or 1 or because the members changed at some point. The detailed breakdown of the repositories into different styles for each year is shown in Table \ref{repo_distribution}. The performance of the Random Forest classifier and the Logistic Regression with recursive feature elimination is shown in Table \ref{team_pred}. As shown in this table, both models had similar performance in predicting the students' teamwork style, Random Forest performed slightly better, with solo-submit being the easiest to predict and collaborative being the most difficult. 

The random forest model worked best with 12 features and the logistic regression worked best with 26. Since random forest performed better at predicting the teamwork style, we analyzed the top features for these random forest models. As these features show, the students' activities in different parts of the project and their prior individual performance were good predictors for their teamwork. Most of the top 12 features selected by random forest for the collaborative, cooperative, and solo-submit classifiers were specific to each of the teamwork style, but some of the features like the \textit{Average deletion per commit for user 0} were common across styles. The \textit{Percentage of commits for the whole project} for both users were the top features for predicting Solo-submitting, while the \textit{Percentage of commits} in different categories and the \textit{Students' prior performance} were more predictive for Collaborative and Cooperative. Surprisingly, the \textit{Number of pair-programming activities} were not among the top features, which might be because the students do not always record pair programming in their commit messages.

\begin{table}[]
    \centering
    \begin{adjustbox}{width=0.48\textwidth}
    \begin{tabular}{|lccc|}
    \hline
    &\multicolumn{3}{c|}{\textbf{Logistic Regression}}\\
     & Collaborative & Cooperative & Solo-submit\\
\hline 
F1 Score & 0.61 & 0.67 & 0.84\\
Precision & 0.51 & 0.70 & 0.90\\
Recall & 0.78 & 0.64 & 0.78\\
\hline 
&\multicolumn{3}{c|}{\textbf{Random Forest}}\\
 & Collaborative & Cooperative & Solo-submit\\
\hline
F1 Score & 0.68 & 0.78 & 0.90\\
Precision & 0.63 & 0.75 & 0.89\\
Recall & 0.79 & 0.83 & 0.92\\
    \hline
    \end{tabular}
    \end{adjustbox}
    \caption{The performance of prediction models for students' teamwork style}
    \label{team_pred}
\end{table}

These prediction models show us that we can identify the students' teamwork style, especially solo-submitting by using automatically generated features from their commit history and their contributions to the different parts of the project. One might assume that looking at the repositories and the students' number of commits should be sufficient for identifying solo-submitters. However, as the SMEs noticed, deciding whether both members had significant contributions to the team was challenging and time-consuming, even for experienced TAs. Most of the defined metrics such as the \textit{Amount of implementation commits} or the \textit{Percentage of commits} in a repository can be extracted  automatically early in the semester. As a result, using predictive models with these features could help identify the need for early intervention, for example when teamwork habits indicate that solo-submit may eventually happen in a team. 

\section{Limitations and Future Work}

There are three main limitations to our work.  First, our dataset is drawn from a single course.  Thus it is possible that the observed results will not generalize to courses with a different team structure or grade breakdown.  We do argue however that the analytical methods we chose are general and we plan to evaluate them on different courses in a future study.  Second, our classification of the student teams was based solely on their observable online behavior and did not consider offline activities. It is possible that offline behaviors such as students meeting face to face, or exchanging code through other media, might affect our results. 

\section{Conclusion}

In this study, we first hypothesized that we could automatically identify students' activities on different parts of development projects based on the text of their commit messages. We later hypothesized in H2 that we could automatically identify different teamwork styles among students using their online submissions and which parts they belong to. For the first part, we manually tagged 400 commit messages as belonging to different parts of the projects as Implementation, Testing, Debug, Style, Documentation, Merge, and Other. We then used TF-IDF features and a logistic regression to automatically label the remaining commits. To analyze different styles in students' teamwork, we manually labeled 100 GitHub repositories of student projects in two offerings of a Java introductory course for CS majors as ``Collaborative'', ``Cooperative'', or ``Solo-submit''. We then used several measures based on the students' activities on GitHub, their prior performance, and whether they chose their teammate to automatically label all the student repositories in these classes. We observed that these models were able to achieve an F1 score of 0.68 or better for different categories, which supported our hypothesis that students' online activities can identify their teamwork style.

The students in these classes were not graded for their amount of contributions on GitHub. As a result, students were able to split the work among themselves based on their choices and what we observed here was their natural behaviors. This makes the findings in this study more likely to apply to other classes since the students' teamwork styles were not directed by the course structure. 

The findings of this study can be used to analyze which styles of teamwork lead to better performance in classes. Eventually, the findings can help design adaptive support platforms for the instructors to observe a summary of the students' activities and possible red flags in their behavior such as solo-submitting. The instructors can then plan interventions in a timely manner to help the students to better engage with authentic team projects in the class.

\section{Acknowledgements}
This research was supported by NSF \#1821475 ``Concert: Coordinating Educational Interactions for Student Engagement'' Collin F. Lynch, Tiffany Barnes, and Sarah Heckman (Co-PIs).
\balance
\clearpage

\bibliographystyle{abbrv}
\bibliography{references}

\end{document}